\documentclass[screen,nonacm,sigconf,natbib=false]{acmart}
\acmYear{2026}
\usepackage[utf8]{inputenc} %
\RequirePackage[
    datamodel=acmdatamodel,
    style=numeric-comp,
    sorting=none,
]{biblatex}
\usepackage[inline]{enumitem} %
\usepackage{siunitx} %
\usepackage{csquotes} %
\usepackage[noEnd]{algpseudocodex} %
\makeatletter
\let\MYcaption\@makecaption
\makeatother
\usepackage{subcaption} %
\makeatletter
\let\@makecaption\MYcaption
\makeatother
\usepackage[capitalize,nameinlink]{cleveref} %
\AtEveryBibitem{%
    \clearname{editor}%
    \clearfield{series}%
    \clearfield{isbn}%
    \clearfield{issn}%
    \clearfield{day}%
    \clearfield{month}%
    \clearfield{place}%
    \clearlist{location}%
    \clearfield{pages}%
    \clearfield{volume}%
    \clearfield{number}%
    \clearfield{eprintclass}%
    \ifentrytype{software}{%
    }{%
        \clearfield{url}%
        \clearfield{urlyear}%
        \clearfield{urldate}%
    }%
}
\DeclareSourcemap{
    \maps[datatype=bibtex]{
        \map{
            \step[fieldsource=doi, match=\regexp{.+/arXiv\..+}, final]
            \step[fieldsource=eprint, match=\regexp{.+}, final]
            \step[fieldset=doi, null]
        }
        \map{
            \step[fieldsource=doi, match=\regexp{.+}, final]
            \step[fieldset=eprint, null]
            \step[fieldset=archiveprefix, null]
            \step[fieldset=eprinttype, null]
        }
    }
}
\setlist[enumerate]{label=(\arabic*)} %
\crefname{section}{Sec.}{Sec.}
\Crefname{section}{Sec.}{Sec.}
\crefname{theorem}{Ex.}{Ex.}
\Crefname{theorem}{Ex.}{Ex.}
\newcommand{\ie}{i.\,e.\@\nobreak}

\newcommand{\eg}{e.\,g.\@\nobreak}

\definecolor{TUM_blue}{RGB}{0,101,189}
\colorlet{TUM_black}{black}
\colorlet{TUM_white}{white}
\definecolor{TUM_darkblue}{RGB}{0,82,147}
\colorlet{TUM_darkblue100}{TUM_darkblue}
\colorlet{TUM_darkblue80}{TUM_darkblue100!80}
\colorlet{TUM_darkblue50}{TUM_darkblue100!50}
\colorlet{TUM_darkblue20}{TUM_darkblue100!20}
\definecolor{TUM_verydarkblue}{RGB}{0,51,89}
\colorlet{TUM_verydarkblue100}{TUM_verydarkblue}
\colorlet{TUM_verydarkblue80}{TUM_verydarkblue100!80}
\colorlet{TUM_verydarkblue50}{TUM_verydarkblue100!50}
\colorlet{TUM_verydarkblue20}{TUM_verydarkblue100!20}
\colorlet{TUM_darkgrey}{TUM_black!80}
\colorlet{TUM_grey}{TUM_black!50}
\colorlet{TUM_lightgrey}{TUM_black!20}
\definecolor{TUM_beige}{RGB}{218,215,203}
\definecolor{TUM_orange}{RGB}{227,114,34}
\definecolor{TUM_green}{RGB}{162,173,0}
\definecolor{TUM_verylightblue}{RGB}{152,198,234}
\definecolor{TUM_lightblue}{RGB}{100,160,200}
\begin{document}
\title[Search Smarter, Not Harder: A Scalable, High-Quality Zoned Neutral Atom Compiler]{Search Smarter, Not Harder:\\A Scalable, High-Quality Zoned Neutral Atom Compiler}
\author{Yannick Stade}
\email{yannick.stade@tum.de}
\orcid{0000-0001-5785-2528}
\affiliation{%
    \institution{Chair for Design Automation, Technical University of Munich}
    \city{Munich}
    \country{Germany}
}
\author{Lukas Burgholzer}
\email{lukas.burgholzer@tum.de}
\orcid{0000-0003-4699-1316}
\affiliation{%
    \institution{Chair for Design Automation, Technical University of Munich}
    \city{Munich}
    \country{Germany}
}
\affiliation{%
    \institution{Munich Quantum Software Company}
    \city{Munich}
    \country{Germany}
}
\author{Robert Wille}
\email{robert.wille@tum.de}
\orcid{0000-0002-4993-7860}
\affiliation{%
    \institution{Chair for Design Automation, Technical University of Munich}
    \city{Munich}
    \country{Germany}
}
\affiliation{%
    \institution{Munich Quantum Software Company}
    \city{Garching near Munich}
    \country{Germany}
}
\hypersetup{ %
    pdftitle={Search Smarter, Not Harder: A Scalable, High-Quality Zoned Neutral Atom Compiler},
    pdfsubject={},
    pdfauthor={
        Yannick Stade,
        Lukas Burgholzer,
        Robert Wille
    }
}
\begin{abstract}
    Zoned neutral atom architectures are emerging as a promising platform for large-scale quantum computing.
    Their growing scale, however, creates a critical need for efficient and automated compilation solutions.
    Yet, existing methods fail to scale to the thousands of qubits these devices promise.
    State-of-the-art compilers, in particular, suffer from immense memory requirements that limit them to small-scale problems.
    This work proposes a scalable compilation strategy that \enquote{searches smarter, not harder}.
    We introduce \emph{Iterative Diving Search}~(IDS), a goal-directed search algorithm that avoids the memory issues of previous methods, and \emph{relaxed routing}, an optimization to mitigate atom rearrangement overhead.
    Our evaluation confirms that this approach compiles circuits with thousands of qubits and, in addition, even reduces rearrangement overhead by \(28.1\%\) on average.
    The complete code is publicly available in open-source as part of the \emph{Munich Quantum Toolkit}~(MQT) at \url{https://github.com/munich-quantum-toolkit/qmap}.
\end{abstract}
\maketitle

\section{Introduction}\label{sec:introduction}

Quantum computing based on neutral atoms~\cite{bluvsteinArchitecturalMechanismsUniversal2025} is reaching a stage where efficient and scalable compilation strategies become indispensable.
In particular, architectures that are split into multiple spatially separated zones featuring dedicated operations---so-called \emph{zoned neutral atom architectures}~\mbox{\cite{bluvsteinArchitecturalMechanismsUniversal2025,bluvsteinLogicalQuantumProcessor2023}}---have been demonstrated as promising candidates for large-scale quantum computing with several thousand qubits~\cite{chiuContinuousOperationCoherent2025}.
These qubit counts render manual compilation strategies or adhoc Python scripts infeasible.
Hence, there is a pressing need for advanced quantum compilation strategies that can efficiently transform high-level quantum algorithms into low-level instructions tailored for this specific architecture.

Unfortunately, despite initial accomplishments in compilation for zoned neutral atom architectures~\cite{stadeAbstractModelEfficient2024,linReuseAwareCompilationZoned2024,stadeRoutingAwarePlacementZoned2025arxiv,ruanPowerMoveOptimizingCompilation2024}, existing approaches either fail to scale to the available qubit counts or compromise on result quality.
In particular, the state-of-the-art compiler~\cite{stadeRoutingAwarePlacementZoned2025arxiv} utilizing \emph{routing-aware placement} improves significantly over previous approaches, \eg, ZAC~\cite{linReuseAwareCompilationZoned2024}.
However, due to immense memory usage, it struggles to compile circuits with more than a few hundreds of qubits and more than a few dozens of parallel entangling gates.
Hence, the question remains how to develop a scalable \emph{and} \mbox{high-quality} compilation strategy.

Instead of searching \emph{harder} by using tens of gigabytes, this work proposes an alternative heuristic compilation strategy.
It combines the established principles of A* search with a much \emph{smarter} exploration strategy.
To this end, we analyze the key aspects in existing compilers allowing them to achieve high-quality results, \ie, low rearrangement overhead.
Based on this analysis, we combine these aspects with a novel search algorithm, called \emph{Iterative Diving Search}~(IDS), that overcomes the scalability issues in previous work---unlocking compilation for large-scale architectures.
To mitigate the increasing rearrangement overhead in large-scale architectures, we propose \emph{relaxed routing}, an optimization that leverages a common conflict-avoidance strategy reducing rearrangement overhead specifically for long distances.
Overall, this eventually leads to a solution that easily scales to thousands of qubits and hundreds of parallel entangling gates while even improving result quality.

Evaluations show that the resulting approach allows, for the first time, to consider instances with up to \SI{5000}{} qubits and \SI{300}{} parallel entangling gates while maintaining high compilation quality.
It compiles all considered circuits within minutes, while the \mbox{state-of-the-art} approach frequently runs into memory outs.
Moreover, it even further improves the compilation quality, \ie, the rearrangement overhead by \(27.1\%\) on average compared to the \mbox{state-of-the-art} method.
Furthermore, the evaluations also unveil an effect of the relaxed routing optimization yielding an overall improvement of \(28.1\%\) on average compared to the \mbox{state-of-the-art}.
The implementation of the proposed approach is publicly available as part of the \emph{Munich Quantum Toolkit}~(MQT,~\cite{willeMQTHandbookSummary2024}) at \url{https://github.com/munich-quantum-toolkit/qmap}.

\section{Background}\label{sec:background}

To provide the necessary context for the compilation problem in zoned neutral atom architectures, this section reviews their computational capabilities and corresponding compilation strategies.

\subsection{Zoned Neutral Atom Architectures}\label{subsec:zoned neutral atom architecture}

\begin{figure}[t]
    \centering
    \includegraphics[width=\linewidth,trim=0 12pt 0 0]{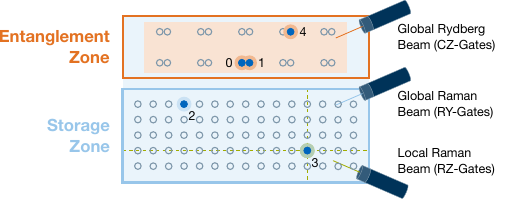}
    \caption{
        Schematic drawing of a zoned neutral atom architecture featuring an entanglement (orange) and a storage zone (light blue).
        Local and global laser beams execute gates.
    }
    \label{fig:zones}
    \vspace{-6pt}
\end{figure}

In quantum computing based on neutral atoms, qubits are encoded in the electronic states of individual neutral atoms from the group of alkali or alkaline-earth metals.
The atoms are trapped by optical tweezers or lattices~\cite{barredoAtombyatomAssemblerDefectfree2016}.
Architectures typically feature one or multiple rectangular grids of such optical traps representing potential atom locations~\cite{bluvsteinLogicalQuantumProcessor2023} depicted as gray circles in~\cref{fig:zones}.

Based on that, operations are realized as follows: single-qubit operations, such as rotations on these atoms, are applied through laser-driven electronic state transitions~\cite{bluvsteinQuantumProcessorBased2022}.
Two-qubit operations such as the controlled phase gate (CZ gate), the common native entangling operation on neutral atom architectures, are realized by the \emph{Rydberg blockade} mechanism~\cite{everedHighfidelityParallelEntangling2023,bluvsteinLogicalQuantumProcessor2023}.
When illuminated with the so-called Rydberg beam, it ensures that only qubits within a certain \emph{interaction radius} of each other interact, while all illuminated atoms, also isolated ones, experience a certain likelihood of an error.

These operations can either be applied locally on individual atoms or globally on all atoms within a defined extent, referred to as a \emph{zone}~\cite{bluvsteinLogicalQuantumProcessor2023}.
In particular, when illuminating the entire entanglement zone with a Rydberg laser, all atoms within that zone are excited to the Rydberg state, but only those pairs of atoms (atom 0 and 1 in \cref{fig:zones}) that are close enough undergo a CZ gate; the others (atom 4) undergo an \mbox{error-prone} identity operation.
To improve the overall fidelity of the computation, zoned architectures spatially separate operations into three different kinds of zones: entanglement zones for two-qubit operations, storage zones featuring long coherence times, and measurement zones for readout~\cite{bluvsteinLogicalQuantumProcessor2023,bluvsteinArchitecturalMechanismsUniversal2025}.

\begin{figure}[t]
    \centering
    \includegraphics[width=\linewidth]{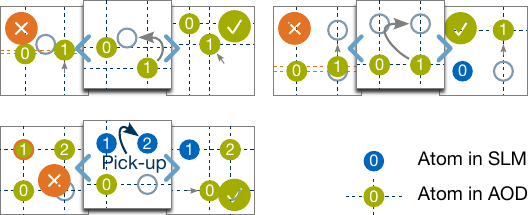}\\
    \raggedright\vspace{-60pt}
    \begin{subfigure}{118pt}
        \caption{Non-crossing constraint}
        \label{fig:constraint:non-crossing}
    \end{subfigure}
    \hfill
    \begin{subfigure}{118pt}
        \caption{Preservation constraint}
        \label{fig:constraint:preservation}
    \end{subfigure}\\
    \vspace{40pt}
    \begin{subfigure}{118pt}
        \caption{Ghost-spot constraint}
        \label{fig:constraint:ghost-spots}
    \end{subfigure}
    \vspace{-8pt}
    \caption{
        The middle frame of each sub-figure shows the intended movement, the left one a violation, and the right one a possible workaround.
    }
    \label{fig:aod constraints}
    \vspace{-12pt}
\end{figure}

The execution of a quantum circuit on a zoned neutral atom architecture involves the following steps:
\begin{enumerate*}
    \item perform all currently executable single-qubit operations,
    \item move pairs of atoms that should interact to the entanglement zone,
    \item perform all currently executable two-qubit operations,
    \item move not needed atoms back to the storage zone,
    \item if applicable, move atoms to the measurement zone for measurements, and
    \item repeat until the entire circuit is executed.
\end{enumerate*}
The movements are achieved by two types of traps: static \emph{Spatial Light Modulator}~(SLM) traps that hold the atoms in place and adjustable \emph{Acousto-Optic Deflector}~(AOD) traps.
Two orthogonal AODs create a rectangular grid of adjustable traps where each row and column of the grid can be controlled individually.
To move an atom, its corresponding column and row are activated to pick it up, shifted to the new position, and then deactivated to drop it off~\cite{bluvsteinQuantumProcessorBased2022}.
While multiple atoms can be moved in parallel, the following \emph{rearrangement constraints} must be followed~\cite{stadeAbstractModelEfficient2024}:
\begin{enumerate}
    \item Columns/rows must maintain a minimal separation; in particular, they cannot cross (cf.~\cref{fig:constraint:non-crossing}).
    \item Activated columns/rows cannot split or merge (cf.~\cref{fig:constraint:preservation}).
    \item Unwanted, so-called \emph{ghost-spots}, must be avoided (cf.~\cref{fig:constraint:ghost-spots}).
\end{enumerate}
One so-called \emph{rearrangement step} starts with picking up some atoms and lasts until dropping off the last atoms.

\subsection{Related Work}\label{subsec:related work}

Compilation for various neutral atom architectures has been an active research area in recent years.
Some of them employ a heuristic approach~\mbox{\cite{bakerExploitingLongDistanceInteractions2021,patelGeyserCompilationFramework2022,nottinghamDecomposingRoutingQuantum2023,patelGRAPHINEEnhancedNeutral2023,schmidHybridCircuitMapping2024,wangAtomiqueQuantumCompiler2024,huangDasAtomDivideandShuttleAtom2024,silverQomposeTechniqueSelect2024,wangQPilotFieldProgrammable2024,ludmirPARALLAXCompilerNeutral2024,tanCompilationDynamicallyFieldProgrammable2025}}, while others use exact methods~\mbox{\cite{constantinidesOptimalRoutingProtocols2024,tanCompilingQuantumCircuits2024,stadeOptimalStatePreparation2024}}.

\begin{figure}[t]
    \centering
    \includegraphics[width=\linewidth]{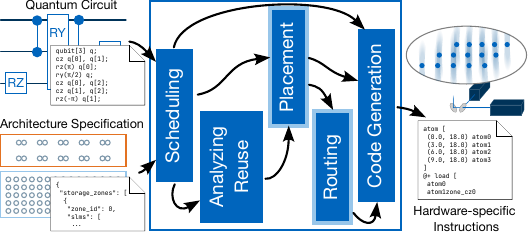}
    \caption{Compilation flow from the quantum circuit and architecture specification to hardware-specific instructions.}
    \label{fig:compilation flow}
    \vspace{-12pt}
\end{figure}

For zoned neutral atom architectures in particular, the following compilation flow as illustrated in \cref{fig:compilation flow} has been established.
It typically consists of the following five steps~\cite{stadeAbstractModelEfficient2024,linReuseAwareCompilationZoned2024,stadeRoutingAwarePlacementZoned2025arxiv,ruanPowerMoveOptimizingCompilation2024}:
\begin{enumerate}
    \item \emph{Scheduling}: divides the circuit into layers of single- and parallel executable two-qubit operations.
    \item \emph{Reuse Analysis}, proposed in~\cite{linReuseAwareCompilationZoned2024}, reduces rearrangement overhead by identifying atoms that can remain in the entanglement zone across multiple layers.
    \item \emph{Placement} assigns atoms, for each layer, to locations.
    \item \emph{Routing} groups movements into rearrangement steps.
    \item \emph{Code Generation} produces hardware-specific instructions.
\end{enumerate}

The first compiler for zoned neutral atom architectures was NALAC~\cite{stadeAbstractModelEfficient2024}.
This compiler, however, also excites idling atoms in the entanglement zone leading to avoidable errors.
To this end, ZAC~\cite{linReuseAwareCompilationZoned2024} improves upon NALAC by moving back all atoms to the storage zone when not needed anymore.
Additionally, it introduces \mbox{\emph{reuse-aware}} compilation to further improve the overall quantum circuit fidelity.
However, ZAC employs a fast but suboptimal placement strategy based on minimum weight perfect matching.
It purely focuses on minimizing the atoms' travel distances, leading to many conflicts with the rearrangement constraints (cf. \cref{subsec:zoned neutral atom architecture}) in the subsequent routing stage.
Due to these conflicts, many movements must be serialized causing many rearrangement steps and resulting in high rearrangement overhead.
PowerMove~\cite{ruanPowerMoveOptimizingCompilation2024} follows a similar approach with the same drawbacks.
These shortcomings are addressed by the \emph{routing-aware placement}~\cite{stadeRoutingAwarePlacementZoned2025arxiv} constituting the current \mbox{state-of-the-art} in compilation for zoned neutral atom architectures.
By considering the rearrangement constraints already during placement, it significantly reduces the number of required rearrangement steps.
Despite this improvement, the employed A* algorithm suffers from severe scalability issues.
In particular, the number of parallel entangling gates in the circuit is crucial for the memory consumption of this compilation strategy.
Compiling circuits with just a few dozen parallel CZ gates per layer can demand over \SI{40}{\giga\byte} of memory, rendering it impractical for larger instances.

\section[Scalable Compilation: Search Smarter, Not Harder]{Scalable Compilation:\\Search Smarter, Not Harder}\label{sec:motivation}

This work introduces a compilation strategy that \enquote{searches smarter, not harder} to overcome the aforementioned limitations.
To this end, we first identify the key features of existing methods responsible for \mbox{high-quality} results and then incorporate them into a novel, more goal-directed search algorithm.
This approach aims to preserve, and even enhance, the result quality while ensuring scalability to thousands of qubits and hundreds of parallel entangling gates.

\subsection{Status Quo: What Makes High-Quality Compilation?}\label{subsec:considered problem}

To mitigate errors caused by decoherence, minimizing the overall computation time of a quantum computation is crucial.
On zoned neutral atom architectures, this translates directly into minimizing the \emph{rearrangement overhead}, \ie, the time spent rearranging atoms between layers.
Since the number of atoms that must be rearranged is given by the circuit's structure, the key determinant to minimize this overhead is the number of atoms that can be rearranged in parallel in a single rearrangement step.
All the individual movements in one rearrangement step must comply with the rearrangement constraints outlined in \cref{subsec:zoned neutral atom architecture}.
Hence, low rearrangement overhead can be achieved by choosing clever placements of atoms for each layer that facilitate rearrangements with many parallel movements.

To this end, the placement stage in \cref{fig:compilation flow} can be modeled as a tree search problem where each node represents a (partial) placement of atoms.
The root node represents the initial placement before any atoms have been placed for the current layer.
To simplify the process, only atoms that need to move, \ie, to or from the entanglement zone are considered.
Each child node extends the placement of its parent node by placing one more additional atom.
A node is a goal node if all atoms have been placed.

To differentiate between good and bad placements, a cost and heuristic function is defined.
The cost function assigns a cost to each node proportional to the rearrangement overhead incurred by the partial placement represented by that node.
The heuristic estimates the remaining cost to reach a goal node, \ie, a complete placement, from the current node.
These functions can be used by heuristic search algorithms, such as the A* algorithm~\cite{hartFormalBasisHeuristic1968}.

The state-of-the-art method demonstrates that a well-designed heuristic and cost function can effectively identify placements that minimize rearrangement overhead~\cite{stadeRoutingAwarePlacementZoned2025arxiv}.
Unlike previous solutions, this method prioritizes placements that preserve the atoms' relative positioning during rearrangement, even at the cost of slightly longer travel distances.
This approach reduces conflicts with rearrangement constraints, enabling more parallel movements per step.
Consequently, a key factor for a successful compilation strategy is a heuristic and cost function that successfully encodes the rearrangement constraints to favor placements that maximize parallelism.
However, for an efficient solution, these functions must be combined with a scalable search strategy.

\begin{figure}[t]
    \centering
    \includegraphics[width=.9\linewidth,trim=0 12pt 0 0]{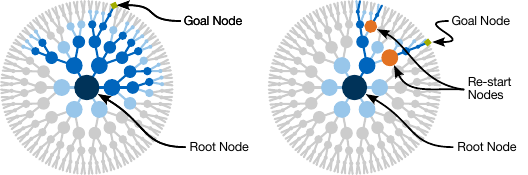}\\
    \raggedright%
    \begin{subfigure}[t]{.48\linewidth}
        \caption{Despite the focus towards goal nodes, the A* algorithm explores the tree more in breath and visits a lot of nodes.}
        \label{subfig:a star}
    \end{subfigure}%
    \hfill%
    \begin{subfigure}[t]{.48\linewidth}
        \caption{The depth-first search-based IDS algorithm explores the tree more streered towards the goal nodes, resulting in fewer nodes being visited.}
        \label{subfig:ids}
    \end{subfigure}\vspace{-8pt}\\
    \caption{When searching for a path from the root node (middle) to a good goal node (green), both algorithms explore nodes (highlighted in blue; the expanded ones in dark blue).}
    \label{fig:graph exploration}
    \vspace{-12pt}
\end{figure}

\begin{example}\label{exp:a star}
The \mbox{state-of-the-art} routing-aware placement~\cite{stadeRoutingAwarePlacementZoned2025arxiv} employs the A* algorithm~\cite{hartFormalBasisHeuristic1968} to search for a low-cost goal node in the search tree.
This leads to a graph exploration as schematically illustrated in \cref{subfig:a star} starting from the middle node marked in green and ending at a (solid green) goal node on the outer circle.
As the illustration shows, the A* algorithm explores a lot of nodes in breath---perfect for an exhaustive search potentially yielding a high-quality result.
However, this also leads to immense memory consumption as all these nodes need to be stored in memory for the A* algorithm.
By that, the search becomes harder and harder as the search space grows, ultimately limiting this approach to only a few dozens of parallel CZ gates per layer in practice.
\end{example}

\subsection{Quo Vadis? A More Goal-Directed Search}\label{subsec:proposed solution}

The \mbox{state-of-the-art} compilation strategy quickly requires more than \SI{40}{GB} of memory when compiling layers with more than a few dozens parallel CZ gates, rendering it impractical for larger instances.
To overcome the memory issue, we propose a \emph{smarter} search strategy that avoids the \emph{hard} job of a very exhaustive search while still maintaining high-quality results.
To this end, we propose \emph{Iterative Diving Search}~(IDS)---a novel search strategy still based on the A* algorithm but exploring the search tree in a more depth-first manner as illustrated in \cref{subfig:ids}.

\begin{example}\label{exp:ids}
As in \cref{exp:a star}, IDS also starts from the initial node in the middle marked green but then dives straight down the tree towards a goal node or a leaf (\ie, node without children).
This behavior resembles a \emph{Depth-First Search}~(DFS) with the important aspect that the most promising child node, according to the heuristic, is always selected for expansion.
After reaching a goal node or a leaf, instead of backtracking like DFS, it jumps to the most promising node discovered so far and restarts the search from there.
In \cref{subfig:ids}, the search is restarted twice, indicated by the nodes marked in orange.
This process is repeated until a constant, configurable number of goal nodes have been found and the best goal node discovered so far is returned.
\end{example}

This more effective search strategy is combined with the cost and heuristic functions that already have proven successful in previous work~\cite{stadeRoutingAwarePlacementZoned2025arxiv} but could easily be updated with upcoming functions.

\subsection{Final Touches: Making a Virtue out of Necessity}\label{subsec:relaxed routing}

Picking up atoms from multiple rows and columns simultaneously can create unwanted ghost-spots, which inadvertently trap atoms that should remain in place.
A common strategy in existing compilers to avoid these ghost-spots is to pick up the atoms row-by-row with a small offset move in between~\cite{linReuseAwareCompilationZoned2024,stadeRoutingAwarePlacementZoned2025arxiv}.
Still, they leave the order of rows and columns unchanged during one rearrangement step.

\begin{figure}[t]
    \begin{subfigure}[t]{\linewidth}
        \centering
        \includegraphics[page=2,width=.85\linewidth]{relaxed_routing}\\
        \vspace{-2pt}%
        \caption{Strict routing requires five rearrangment steps (the last step is omitted for brevity), as described in~\cref{exp:strict routing}.}
        \label{subfig:strict routing}
    \end{subfigure}\vspace{8pt}\\
    \begin{subfigure}[t]{\linewidth}
        \centering
        \includegraphics[page=1,width=.85\linewidth]{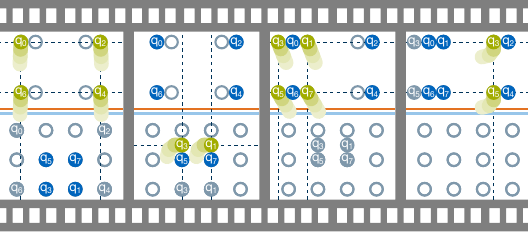}\\
        \vspace{-2pt}%
        \caption{Relaxed routing accomplishes the entire rearrangement in only two steps by shifting the lower row upwards before picking up the next row and allowing the left column to shift right after dropping off the right column.}
        \label{subfig:relaxed routing}
    \end{subfigure}
    \vspace{-8pt}
    \caption{
        Comparison of routing approaches for an interaction of the atom pairs \((0, 1), (2, 3), (4, 5)\) and \((6, 7)\).
    }
    \label{fig:routing approaches}
    \vspace{-12pt}
\end{figure}

\begin{example}\label{exp:strict routing}
    In \cref{fig:routing approaches}, a CZ gate is to be performed between the atom pairs \((0, 1)\), \((2, 3)\), \((4, 5)\), and \((6, 7)\).
    To this end, the atoms need to be aligned accordingly in the entanglement zone (above the orange line).
    Using strict routing, four atoms can be moved in one rearrangement step as shown in the first frame of \cref{subfig:strict routing}.
    However, to avoid ghost-spots, strict routing requires four more separate rearrangement steps to move the remaining atoms.
\end{example}

One can make a virtue out of this necessity to avoid ghost-spots by dropping the \emph{strict} constraint that the rows must remain in order during a rearrangement step.
This constraint can be \emph{relaxed} by allowing already activated rows to shift vertically before the next row is picked up.
Analogously, when allowing to drop off atoms column-by-column (instead of row-by-row) and shifting still activated columns horizontally after dropping off a column, also the columns can be reordered.
We call this new approach \emph{relaxed routing} in contrast to the established \emph{strict routing}.

\begin{example}\label{exp:relaxed routing}
Relaxed routing, as shown in \cref{subfig:relaxed routing}, picks up atom~\(1\) and \(3\) (2nd frame), shifts them upwards, and then picks up atom~\(5\) and \(7\) (3rd frame)---effectively reordering the rows.
Similarly, after dropping off atom~\(1\) and \(7\) (4th frame), the left column is shifted right after the first drop-off---effectively reordering the columns.
\end{example}

\section{Implementation}\label{sec:implementation}

The realization of the proposed methods requires careful implementation.
The following sections outline the details on the implementation of IDS and the tuning of relaxed routing.

\subsection{Iterative Diving Search}\label{subsec:ids implementation}

The following pseudocode implements the proposed \emph{Iterative Diving Search}~(IDS).
Without the parts highlighted in blue, the pseudocode implements the classical A* algorithm~\cite{hartFormalBasisHeuristic1968}.\footnote{Note that the search space resembles a tree, and hence there is no need to remember visited nodes because there are no cycles.}\vspace{2pt}

\noindent\includegraphics[width=\linewidth]{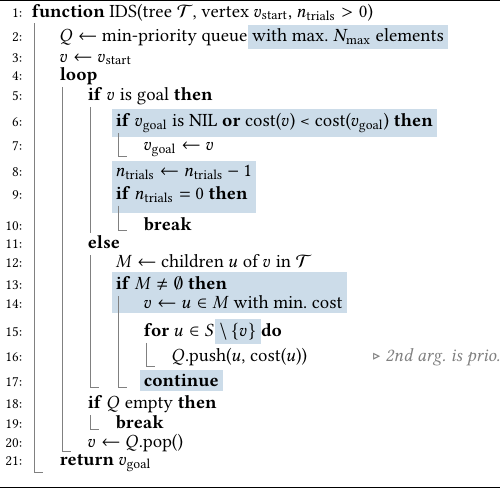}

IDS uses a min-priority queue \(Q\) to keep track of the most promising nodes encountered so far that have not yet been expanded (line~2).
To bound the memory consumption, the maximum number of elements \(N_\mathrm{max}\) in \(Q\) is limited.
This may cause some nodes to be discarded when \(Q\) is full, leading to a non-exhaustive search.
When a new goal node is found, the best goal node discovered so far is updated (lines~\mbox{6--7}).
Additionally, the variable \(n_\mathrm{trials}\)---used to keep track of how many goal nodes should be found before terminating the search---is decremented accordingly (lines~\mbox{8--9}).

While expanding a node with children, the child with the lowest cost is \emph{not} added to \(Q\) and directly selected for expansion (lines~\mbox{13--14}).
This implements the depth-first exploration of IDS.
If the node is a leaf, the most promising node from \(Q\) is selected for expansion (lines~\mbox{18--19})---the same applies after a goal node has been discovered.
The search either ends when \(n_\mathrm{trials}\) reaches zero or when no more nodes are available for expansion (lines~10 and~19).

\subsection{Tuning Relaxed Routing}\label{subsec:relaxed trade-off}

It is expected that the relaxed routing approach especially pays off for long rearrangement distances.
This is due to the effect that the additional offset moves introduced by relaxed routing also incur some overhead.
For short rearrangements, this additional overhead might even outweigh the benefits of relaxed routing.

\begin{figure}[t]
    \centering
    \includegraphics[page=3,width=.9\linewidth]{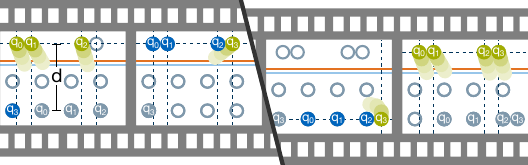}\\
    \raggedright\vspace{-12pt}%
    \begin{subfigure}[t]{0.48\linewidth}
        \caption{Strict Routing}
        \label{subfig:strict option}
    \end{subfigure}%
    \hfill%
    \begin{subfigure}[t]{0.48\linewidth}
        \caption{Relaxed Routing}
        \label{subfig:relaxed option}
    \end{subfigure}\vspace{-8pt}\\
    \caption{
        \textbf{(a)}~To let the atom pairs \((0,1)\) and \((2,3)\) interact, the strict routing approach requires two separate rearrangement steps.
        \textbf{(b)}~By first picking up atom \(3\), moving it right of atom \(2\), and then picking up the remaining atoms, the relaxed routing can achieve the same rearrangement in one go.
        The cost for the additional offset move particularily pays off when the distance \(\mathsf{d}\) indicated in \cref{subfig:strict option} is large.
    }
    \label{fig:hybrid routing}
    \vspace{-12pt}
\end{figure}

\begin{example}\label{exp:relaxed trade-off}
    \Cref{fig:hybrid routing} shows the rearrangements to execute CZ gates between the atom pairs \((0,1)\) and \((2,3)\) using strict routing in \cref{subfig:strict option}, and relaxed routing in \cref{subfig:relaxed option}.
    When the distance \(\mathsf{d}\) is small, the offset move of atom \(3\) in \cref{subfig:relaxed option} incurs a higher time overhead than the extra rearrangement step required by strict routing in \cref{subfig:strict option}.
    However, as \(\mathsf{d}\) increases, the time overhead of the extra rearrangement step outweighs the overhead of the offset movement.
\end{example}

Hence, there exists a crossover point depending on the rearrangement distance where relaxed routing starts to pay off.
To this end, we estimate the additional overhead caused by relaxed routing and fall back to strict routing if preferable.

\section{Evaluation}\label{sec:evaluation}

The proposed strategy is the first one that combines high-quality results and scalability.
To evaluate the effectiveness of the proposed methods, we compare its performance against the \mbox{state-of-the-art} compilation strategy for zoned neutral atom architectures~\cite{stadeRoutingAwarePlacementZoned2025arxiv}.
This section starts with a review of the experimental setup and a parameter study, followed by a presentation and discussion of the obtained results.
The complete code is publicly available in open-source as part of the MQT~\cite{willeMQTHandbookSummary2024} under \url{https://github.com/cda-tum/mqt-qmap/}.

\subsection{Experimental Setup and Parameter Study}\label{subsec:setting}

We implemented the proposed method in C++ on top of the existing \mbox{state-of-the-art} compiler~\cite{stadeRoutingAwarePlacementZoned2025arxiv}.
All experiments were conducted on an Apple M3 with 16GB of RAM.
During compilation, a memory limit of \SI{20}{\giga\byte} was enforced for each instance.

We considered the zoned neutral atom architecture proposed in~\cite{bluvsteinLogicalQuantumProcessor2023} slightly enlarged to an overall dimension of \SI{400}{} by \SI{400}{\micro\meter} to fit the benchmarks.
The architecture is composed of two zones: a storage zone with a \(73 \times 101\) grid of traps separated by \SI{4}{\micro\meter}, and an entanglement zone with a \(10 \times 34\) grid of trap pairs.
Each pair has two traps positioned (horizontally) \SI{2}{\micro\meter} apart, and the pairs are spaced to maintain a minimum distance of \SI{10}{\micro\meter} between traps from adjacent pairs.
The zones are vertically arranged, horizontally centered, and have a minimum separation of \SI{21}{\micro\meter}.

To benchmark the proposed approach, we have selected various quantum circuits of varying sizes from MQT Bench~\cite{quetschlichMQTBenchBenchmarking2023}, as listed in the first two columns of~\cref{tab:results ids}.
The next four columns list key characteristics of the circuits; in particular, the maximum number of parallel two-qubit gates per layer being the most critical factor for the size of the search space, as discussed in \cref{subsec:related work}.
Due to their structure, most of these circuits yield a low number of parallel entangling gates per layer.
One exception is the \texttt{graphstate} family, where the degree of parallelism grows with the circuit size until the capacity of the architecture's entanglement zone is reached.\footnote{The architecure's limit is 306 parallel CZ gates as the compiler employs a 90\% filling at maximum.}
We consider these highly parallel circuits representative of circuits specifically tailored for neutral atom architectures, which offer more parallelism than, for instance, superconducting architectures.

\begin{figure}[t]
    \centering
    \includegraphics[width=\linewidth,trim=0 24pt 0 0]{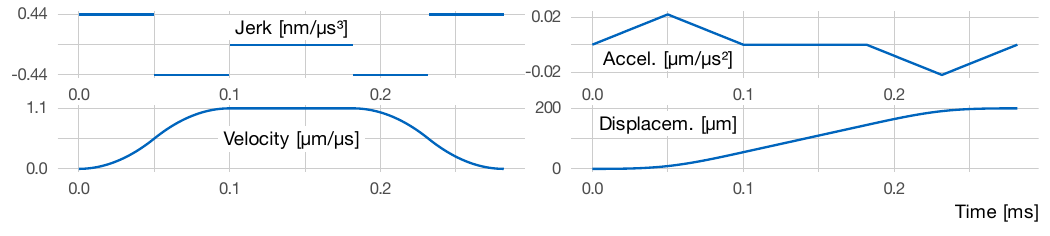}
    \caption{Movement profile with a constant jerk and a maximum speed.}
    \label{fig:movement profile}
    \vspace{-12pt}
\end{figure}

As the central metric, to evaluate the compilation quality, we measure the time overhead caused by rearrangements during the execution of the compiled circuit, referred to as \emph{rearrangement time}.
This metric directly impacts the overall fidelity of the computation because it contributes to the total computation time during which decoherence can occur.
We assume that pick-up and drop-off operations take \SI{15}{\micro\second}~\cite{bluvsteinLogicalQuantumProcessor2023,linReuseAwareCompilationZoned2024} and employ a movement profile based on a constant jerk of \SI{0.44}{\nano\meter\per\second\cubed} with a maximum speed of \SI{1.1}{\micro\meter\per\micro\second}~\cite{bluvsteinLogicalQuantumProcessor2023}.
\Cref{fig:movement profile} shows the resulting movement profile for a distance of \SI{200}{\micro\meter}.
The resulting time for this movement is \SI{0.24}{\milli\second}.

\begin{figure}[t]
    \centering
    \includegraphics[width=\linewidth,trim=0 12pt 0 0]{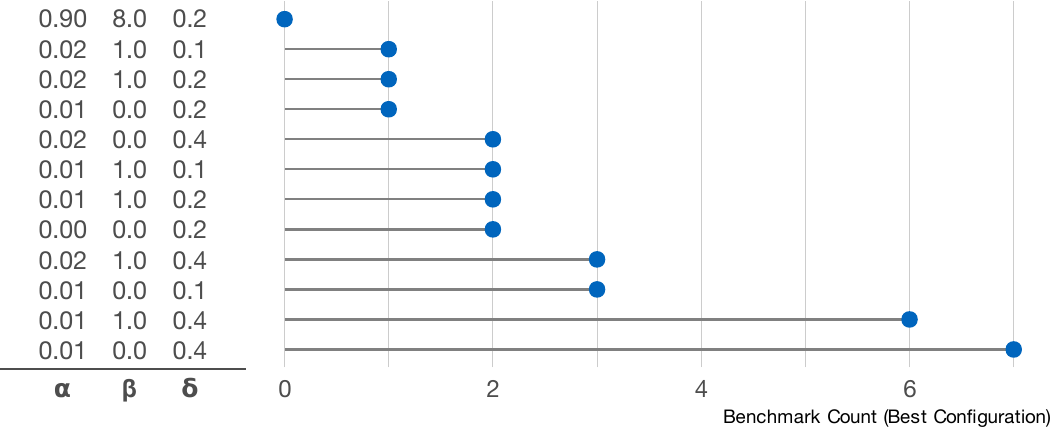}
    \caption{The plot counts the benchmarks (x-axis) for which a specific configuration (y-axis) yielded the best result.}
    \label{fig:parameter study}
    \vspace{-12pt}
\end{figure}

As explained in \cref{subsec:proposed solution}, the \mbox{state-of-the-art} and the proposed method employ the same tunable heuristic function.
For the \mbox{state-of-the-art} approach, we use the best parameter configuration identified for large circuits in~\cite{stadeRoutingAwarePlacementZoned2025arxiv}.
However, since our experiments indicated this configuration is suboptimal for the proposed method, we conducted a parameter study to find a better-suited configuration for the proposed approach.
To this end, we varied the parameters \(\delta, \beta\) (accelerating part of the heuristic), and \(\alpha\) (\mbox{look-ahead}, cf.~\cite{stadeRoutingAwarePlacementZoned2025arxiv}) and summarized the results in \cref{fig:parameter study}.
The best results were achieved with the configuration \mbox{\(\delta = 0.01\)}, \mbox{\(\beta = 0.0\)}, and \mbox{\(\alpha = 0.4\)}.

\subsection{Obtained Results}\label{subsec:eval ids}

\begin{table*}[t]
    \setlength{\aboverulesep}{0.4pt} %
    \setlength{\belowrulesep}{0.4pt} %
    \caption{Results of Iterative Diving Search}\vspace{-10pt}%
    \label{tab:results ids}
    \begin{tabular}{%
        p{58pt}%
        S[table-format=4.0]%
        S[table-format=5.0]%
        S[table-format=4.0]%
        S[table-format=3.0]%
        |S[table-format=2.1]%
        S[table-format=4.1]%
        |S[table-format=3.1]%
        @{\ (\hspace{.2ex}}r@{)\hspace{2ex}} %
        S[table-format=5.1]%
        @{\ (\hspace{.2ex}}r@{)\hspace{2ex}} %
        |S[table-format=5.1]%
        @{\ (\hspace{.2ex}}r@{)\hspace{2ex}} %
    }
        \toprule
        \multicolumn{4}{l}{\textbf{Benchmark}} & {Max.} & \multicolumn{2}{c|}{\textbf{A*} \textit{(State-of-the-Art)}} & \multicolumn{4}{c|}{\textbf{IDS} \textit{(Proposed Solution)}} & \multicolumn{2}{c}{\textbf{Relaxed Routing}} \\[-1pt]
        \cmidrule{6-7}\cmidrule{8-11}
        \multicolumn{2}{r}{Num.} & {Num.} & {Num.} & {\hspace{-3pt}2Q-Gates\hspace{-3pt}} & {Comp.} & {Rearr.} & \multicolumn{2}{c}{Comp.} & \multicolumn{2}{c|}{Rearr.} & \multicolumn{2}{c}{\textit{(Proposed Opt.)}} \\[-1pt]
        \cmidrule{12-13}
        \multicolumn{2}{r}{Qubits} & {2Q-Gates} & {Layers} & {\hspace{-3pt}in Layer\hspace{-3pt}} & {Time [\si{\second}]} & {T. [\si{\milli\second}]} & \multicolumn{2}{c}{Time [\si{\second}]} & \multicolumn{2}{c|}{Time [\si{\milli\second}]} & \multicolumn{2}{c}{Rearr. Time [\si{\milli\second}]} \\
        \midrule
                  qft &  500 & 18620 &  1994 &  10 &                4.0 & 2786.3 &   4.2 &   $+$0.2 & 2324.5 & $-$16.6\% & 2287.1 & $-$17.9\% \\
                      & 1000 & 37620 &  3994 &  10 &                4.9 & 5816.2 &   6.2 &   $+$1.3 & 4937.9 & $-$15.1\% & 4855.8 & $-$16.5\% \\
               wstate &  500 &   998 &   501 &   2 &                0.5 &  421.1 &   0.6 &   $+$0.1 &  410.8 &  $-$2.4\% &  410.5 &  $-$2.5\% \\
                      & 1000 &  1998 &  1001 &   2 &                1.0 &  945.0 &   1.0 & $\pm$0.0 &  777.7 & $-$17.7\% &  777.1 & $-$17.8\% \\
             qpeexact &  500 & 19579 &  2987 &  10 &                5.3 & 3316.4 &   5.3 & $\pm$0.0 & 2867.5 & $-$13.5\% & 2825.2 & $-$14.8\% \\
                      & 1000 & 39579 &  5987 &  10 & \multicolumn{2}{c|}{MEMOUT} &   6.7 &     {NA} & 5928.0 &      {NA} & 5845.1 &      {NA} \\
        vqe two local &   50 &  3675 &   197 &  25 &                2.7 &  806.5 &   2.7 &   $-$0.1 &  631.9 & $-$21.7\% &  624.7 & $-$22.5\% \\
                      &  100 & 14850 &   397 &  50 &               20.5 & 3662.5 &  14.7 &   $-$5.8 & 2423.1 & $-$33.8\% & 2394.6 & $-$34.6\% \\
                      &  150 & 33525 &   597 &  75 &               63.0 & 7994.3 &  47.5 &  $-$15.6 & 4432.0 & $-$44.6\% & 4374.9 & $-$45.3\% \\
                      &  200 & 59700 &   797 & 100 & \multicolumn{2}{c|}{MEMOUT} & 120.1 &     {NA} & 8477.0 &      {NA} & 8349.7 &      {NA} \\
                 qaoa &   50 &  2348 &   248 &  16 &                1.3 &  410.0 &   1.4 & $\pm$0.0 &  270.1 & $-$34.1\% &  267.5 & $-$34.8\% \\
                      &  100 &  9736 &   508 &  31 & \multicolumn{2}{c|}{MEMOUT} &   8.2 &     {NA} &  995.2 &      {NA} &  979.8 &      {NA} \\
                      &  150 & 22256 &   762 &  47 & \multicolumn{2}{c|}{MEMOUT} &  26.7 &     {NA} & 2026.5 &      {NA} & 1996.7 &      {NA} \\
                      &  200 & 39552 &  1028 &  61 & \multicolumn{2}{c|}{MEMOUT} &  67.6 &     {NA} & 3650.8 &      {NA} & 3593.9 &      {NA} \\
           graphstate &   60 &    60 &     6 &  22 &                0.1 &   16.6 &   0.1 & $\pm$0.0 &   14.8 & $-$11.2\% &   14.8 & $-$11.2\% \\
                      &   80 &    80 &     6 &  30 &                0.2 &   31.9 &   0.2 &   $+$0.1 &   19.7 & $-$38.1\% &   19.4 & $-$39.3\% \\
                      &  100 &   100 &    10 &  34 &                0.3 &   40.8 &   0.3 & $\pm$0.0 &   22.4 & $-$45.1\% &   22.1 & $-$45.9\% \\
                      &  120 &   120 &     6 &  43 &               10.5 &   45.0 &   0.5 &  $-$10.0 &   28.7 & $-$36.3\% &   28.4 & $-$36.8\% \\
                      &  140 &   140 &     9 &  48 & \multicolumn{2}{c|}{MEMOUT} &   0.5 &     {NA} &   36.6 &      {NA} &   36.2 &      {NA} \\
                      &  160 &   160 &     6 &  59 & \multicolumn{2}{c|}{MEMOUT} &   1.5 &     {NA} &   38.0 &      {NA} &   37.7 &      {NA} \\
                      &  200 &   200 &     7 &  77 & \multicolumn{2}{c|}{MEMOUT} &   1.2 &     {NA} &   53.0 &      {NA} &   51.8 &      {NA} \\
                      &  500 &   500 &     6 & 184 & \multicolumn{2}{c|}{MEMOUT} &   8.0 &     {NA} &  133.8 &      {NA} &  132.8 &      {NA} \\
                      & 1000 &  1000 &     7 & 306 & \multicolumn{2}{c|}{MEMOUT} &  57.6 &     {NA} &  320.5 &      {NA} &  320.9 &      {NA} \\
                      & 2000 &  2000 &     9 & 306 & \multicolumn{2}{c|}{MEMOUT} & 227.4 &     {NA} &  802.7 &      {NA} &  802.1 &      {NA} \\
                      & 5000 &  5000 &    18 & 306 & \multicolumn{2}{c|}{MEMOUT} & 647.7 &     {NA} & 2588.8 &      {NA} & 2582.3 &      {NA} \\
        \midrule
        \(\varnothing\) &    &       &       &     &                8.8 & 2022.5 &  50.3 &   $-$2.3 & 1768.5 & $-$27.1\% & 1745.2 & $-$28.1\% \\
        \bottomrule
    \end{tabular}
\end{table*}

Using the setup and the best parameter configuration identified in \cref{subsec:setting}, we eventually compare the proposed method against the \mbox{state-of-the-art} method proposed in~\cite{stadeRoutingAwarePlacementZoned2025arxiv}.
\Cref{tab:results ids} summarizes the obtained results across all benchmarks.
The middle two regions of \cref{tab:results ids} list the compilation times and resulting rearrangement times for the \mbox{state-of-the-art} A*-based method and the proposed IDS method, respectively.
The final column shows rearrangement times for IDS combined with relaxed routing.

The results reveal that compilation times of both methods are comparable with the notable exception that the proposed method requires significantly less memory and never hits the memory limit of \SI{20}{\giga\byte}.
This allows the proposed method to successfully compile larger instances while the \mbox{state-of-the-art} method frequently fails due to memory outs.
Comparing the resulting rearrangement times as the most critical quality metric, the proposed method consistently outperforms the \mbox{state-of-the-art} method, achieving an average improvement of \(27.1\%\) and up to \(45.1\%\) on highly parallel benchmarks like the \texttt{graphstate} family.
Adding the proposed relaxed routing further reduces rearrangement overhead---reaching an overall improvement of up to \(45.9\%\) for specific benchmarks and \(28.1\%\) on average compared to the previous \mbox{state-of-the-art}.

These results demonstrate that the proposed method's more goal-directed search strategy finds significantly better placements despite being less exhaustive.
Unlike the A*-based approach, which requires a heuristic balancing solution quality with search acceleration~\cite{stadeRoutingAwarePlacementZoned2025arxiv}, the inherent efficiency of the proposed method allows for a heuristic focused solely on solution quality, resulting in superior placements.
Moreover, relaxed routing can successfully mitigate rearrangement overhead that grows with circuit size.

\section{Conclusions}\label{sec:conclusions}

This work addressed the scalability limitations of existing compilers for zoned neutral atom architectures. %
We introduced a scalable compilation strategy, called \emph{Iterative Diving Search}~(IDS), and a novel optimization called \emph{relaxed routing}.
Evaluations show that this approach not only scales to thousands of qubits but also reduces the rearrangement overhead by \(28.1\%\) on average compared to the state-of-the-art.
These advancements enable efficient compilation of large-scale, highly parallel quantum circuits on near-term neutral atom quantum computers.

\begin{acks}
The authors acknowledge funding from the European Research Council (ERC) under the European Union’s Horizon 2020 research and innovation program (grant agreement No. 101001318), and the Munich Quantum Valley (MQV), which is supported by the Bavarian state government with funds from the Hightech Agenda Bayern Plus.
\end{acks}

\printbibliography

\end{document}
\typeout{get arXiv to do 4 passes: Label(s) may have changed. Rerun}